\documentclass[prl, twocolumn, amsmath,floatfix]{revtex4}

\usepackage{graphicx}
\usepackage{epstopdf}
\usepackage{natbib}

\newcommand{\Alfven}{Alfv\'{e}n}
\newcommand{\sciexp}[2]{{#1}\ensuremath{\,\times\,10^{#2}}}
\newcommand{\pow}[1]{\ensuremath{^{{#1}}}}

\newcommand{\PSC}{\textsc{psc}}

\begin{document}

\begin{abstract}
Recent experiments have observed magnetic reconnection in high-energy-density,
laser-produced plasma bubbles, with reconnection rates observed to
be much higher than can be explained by classical theory.
Based on fully kinetic particle simulations we find that fast reconnection
in these strongly driven systems can be explained by magnetic flux 
pile-up at the shoulder of the current sheet and subsequent
fast reconnection via two-fluid, collisionless mechanisms.
In the strong drive regime with two-fluid effects, we find that the
ultimate reconnection time is insensitive to the nominal
system \Alfven{} time.

\end{abstract}

\title{Fast magnetic reconnection in laser-produced plasma bubbles}
\author{W. Fox, A. Bhattacharjee and K. Germaschewski}
\affiliation{Center for Integrated Computation and Analysis of Reconnection and Turbulence, \\
           and Center for Magnetic Self-Organization in Laboratory and Astrophysical Plasmas, \\
University of New Hampshire, Durham, NH 03824}
\date{\today}

\maketitle

Magnetic reconnection \cite{YamadaRMP2010, Biskamp2000}, the change of magnetic
topology in the presence of plasma, is
observed in space, laboratory, and, most recently, laser-produced
high-energy-density (HED) plasmas 
\cite{NilsonPRL2006, LiPRL2007b, NilsonPoP2008, WillingalePoP2010}.
Reconnection plays a key
role in energy release by plasma instabilities, as in solar
flares or magnetospheric substorms, and the change in topology
allows the rapid heat transport associated with sawtooth relaxation in
magnetic fusion devices \cite{HastieAstroSpaceSci1997}.
The observation of reconnection in HED plasmas suggests
that it may also play a role in inertial confinement fusion, and
indeed recent work has now observed the generation of large-scale magnetic fields
during inertial fusion implosions \cite{RyggScience2008}.

\begin{figure}
\centering
\includegraphics{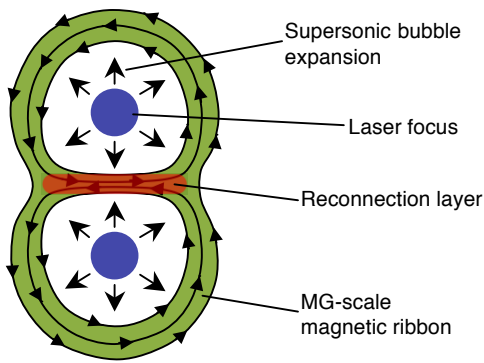}
\caption{Magnetic reconnection driven between expanding plasma bubbles,
view from top.}
\label{FigBubble}
\end{figure}

We investigate recent experimental observations 
\cite{NilsonPRL2006, LiPRL2007b, NilsonPoP2008, WillingalePoP2010}
of fast magnetic reconnection between the
HED plasma bubbles created by focusing 
terawatt (TW)-class lasers ($\sim$kJ/ns) down to 
sub-millimeter-scale spots on a plastic or metal foil. 
The foil is ionized into hemispherical bubbles that expand 
supersonically off the surface of the foil.  Each bubble is found to self-generate a
strong magnetic field of order megagauss (MG), which forms a
toroidal ribbon wrapping around the bubble.  If multiple bubbles 
are created at small separation, the bubbles expand into one 
another, and the opposing magnetic fields are squeezed 
together and seen to reconnect (Fig.~\ref{FigBubble}). 
The rates of reconnection are observed to be fast, and unexplained by classical 
Sweet-Parker theory \cite{NilsonPRL2006, LiPRL2007b}.  
Reconnection has also been observed
between 
laser-produced-plasma bubbles immersed 
in a background plasma and magnetic field \cite{GekelmanPoP2007},
though we focus here on reconnection between HED bubbles
driven by kJ-class lasers.

It is of great interest to bring these results in line with what is already known 
about reconnection.  As mentioned above, there are a number of new 
features in these laser-driven experiments, such as the high energy 
density in the plasma and magnetic field.  Perhaps their 
most notable feature is the very strong reconnection drive: 
the opposing magnetic fields are driven together by the 
expanding bubbles at sonic and super-\Alfven{}ic velocities.
Such a strongly-driven regime has not been previously accessible
in experiments, but may aid understanding of reconnection at the Earth's magnetopause (where the driver of reconnection
is the super-\Alfven{}ic solar wind) and other astrophysical
contexts with high plasma $\beta$ (accretion disks, stellar interiors),
or with colliding supersonic, magnetized flows (supernovae remnants,
or at the heliopause).  At the magnetopause, for example, the strong
inflow can drive ``flux-pileup'' reconnection.
\cite{AndersonJGR1997, MorettoAnnGeo2005}.
%Signatures of flux-pileup reconnection have also appeared
%in simulations of magnetohydrodynamic turbulence \cite{ServidioPRL2009}.

\begin{figure*}[t]
\centering
\includegraphics{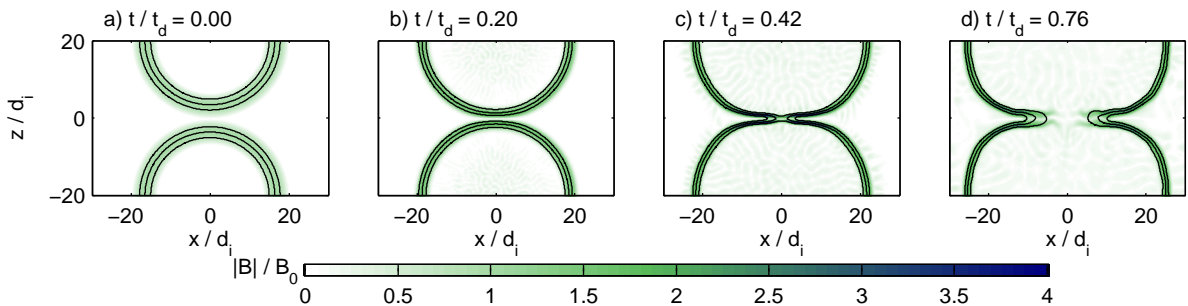}
\caption{Evolution of $|B|$ and contours of magnetic flux
during the simulation.}
\label{FigBevol2d}
\end{figure*}

Here, we report the results of the first fully kinetic, particle-in-cell (PIC) 
simulations with parameters and geometry relevant to these experiments.
Our main aim has been 
to understand the apparent \Alfven{}ic reconnection rates 
observed in the experiments.  We demonstrate that it can be 
explained as a combination of two effects: (1) a strong pile-up 
of magnetic flux \cite{BiskampPoF1986} at the shoulder of the current sheet, 
caused by the sonic inflow and corresponding ram pressure 
that drives substantial compression of the plasma and magnetic 
field between the two bubbles, and (2)
the intervention of collisionless effects 
(Hall current and electron pressure tensor) 
when the current sheet width falls 
below the ion skin depth ($d_i = c/\omega_{pi}$, where $c$ 
is the speed of light, and 
$\omega_{pi}$ the ion plasma frequency).  
Previous discussion of these experiments has proposed anomalous resistivity \cite{NilsonPoP2008},
or the Hall effect \cite{WillingalePoP2010}, but have
not discussed pileup, which we find is an essential ingredient.
To our knowledge, these laser-driven experiments are the first in 
which flux pile-up as well as collisionless effects play important roles, producing
highly time-dependent, impulsive reconnection dynamics, and
\Alfven{}ic peak reconnection rates.

In the laser-driven experiments, the reported reconnection rates
were very fast compared to standard, 
classical predictions from Sweet-Parker theory, 
and instead were consistent with a simple
hydrodynamic time based on the observed bubble expansion rate.
The reconnection inflows (1.3 and \sciexp{4}{5}~m/s
for Rutherford \cite{NilsonPoP2008} and Omega \cite{LiPRL2007b},
respectively), ranged from about 0.7 to 3 times the 
associated \Alfven{} speed. ($V_A = B / \sqrt{\mu_0 \rho}$,
calculated based on $B$ fields (100~T, 50~T), and
mass density $\rho$ derived from ion species (Al, CH) 
and electron density (both \sciexp{$\sim$5}{25}~m\pow{-3}),
evaluated at the bubble edge, before the onset of the bubble collision.)
Despite the difficulty making these novel measurements, 
the overall picture is that the reconnection is very strongly driven
and occurs at \Alfven{}ic to super-\Alfven{}ic rates.
Naturally then, the inflows are also much larger than allowed
by classical Sweet-Parker theory, which allows inflows only
a fraction of the \Alfven{} speed, $V_{SP} \sim V_A / S^{1/2}$,
where the Lundquist number $S = L V_A \mu_0 / \eta$, 
based on system size $L$ and classical resistivity $\eta$,
roughly 200 and 1000 for Rutherford and Omega.
Meanwhile, the ion skin depths $d_i$  (40~$\mu$m, 40~$\mu$m)
are equal or larger than 
Sweet-Parker current sheet widths $\delta_{SP} = L/S^{1/2}$
(15, 30~$\mu$m),
suggesting that the experiments are in the 
regime for two-fluid, collisionless reconnection.
The two-fluid reconnection mechanisms have been extensively 
observed in simulations \cite{MaGRL1996,BirnJGR2001} and
have begun to be verified experimentally \cite{YamadaPoP2006}.
Given this, however, the reported \Alfven{}ic reconnection rates
remain extraordinary, because even these fast, collisionless
theories still typically find reconnection inflows ``only'' near
0.1--0.2~$V_{A}$.

We now present 2-d particle-in-cell simulations
of bubble reconnection.  The simulations 
track a pair of expanding bubbles through their
interaction and reconnection, which is
driven by the substantial energy stored in 
the plasma flow.  Our simulations adopt an initial 
condition corresponding to a time
about halfway through the experiments, after the bubbles are
created, have expanded, and have generated their 
magnetic ribbons, but before the pair of
bubbles has begun to interact.  (It thus avoids the
physics of the magnetic field generation process,  
which is a 3-dimensional, $\nabla n \times \nabla T_e$
two-fluid effect \cite{NilsonPRL2006, LiPRL2007b}.)
We then model the subsequent expansion and interaction
of the plasma bubbles.

We define the following initial conditions.
The system here runs, in $(x,z)$ coordinates,
over the domain $[-L_x, L_x] \times [-L_z, L_z]$.  
The system is periodic in both $x$ and $z$ for 
computational expediency.  It therefore
contains two half-bubbles,
one centered at $(0, -L_z)$ and the other at $(0, +L_z)$.
Define the radius vectors from
the center of each bubble, $\mathbf{r}^{(1)} = (x, z+L_z)$ and
$\mathbf{r}^{(2)} = (x, z-L_z)$.
Then the initial density is $n_b + n^{(1)} + n^{(2)}$,
where $n_b$ is a background density, 
and the densities $n^{(i)}$ are 
\begin{equation}
n^{(i)}(x,z) = 
\begin{cases} 
  (n_0-n_b) \cos^2 \left(\frac{\pi r^{(i)}}{2 L_n}\right) & \mbox{if } r^{(i)} < L_n, \\
   0 & \mbox{otherwise.}
\end{cases}
\end{equation}
Here $L_n$ is the initial size of the bubbles.  Next, 
the velocity field is initialized as the sum of the fields
%$\mathbf{V}^{(1)}$ and $\mathbf{V}^{(2)}$ of two radially expanding bubbles,
\begin{equation}
\mathbf{V}^{(i)} =  
\begin{cases}
   V_0 \sin \left(\frac{\pi r^{(i)}}{L_n}\right) \mathbf{\hat{r}}^{(i)}
    & \mbox{if $r^{(i)} < L_n$,} \\
   0 & \mbox{otherwise.}
\end{cases}
\end{equation}
%\begin{equation}
%\mathbf{V}^{(i)} =  
%   M_0 \left({T_{e0}/M_i}\right)^{1/2} \sin \left(\frac{\pi r^{(i)}}{L_n}\right) \mathbf{\hat{r}}^{(i)},
%\end{equation}
%for $r^{(i)} < L_n$ (i.e. within the bubble), and zero otherwise.
These simulations adopt uniform 
initial electron and ion temperatures ($T_{e0}$ and $T_{i0}$) for simplicity.
The magnetic field is initialized (Fig.~\ref{FigBevol2d}(a)) 
as the sum of two toroidal ribbons, with
%\begin{equation}
%\mathbf{B}^{(i)} =  B_0 \sin \left( \frac{\pi (r^{(i)}-L_n)}{2 L_B} \right) 
%           \mathbf{\hat  {r}}^{(i)}  \times \mathbf{\hat {y}},
%\label{EqBInitial}
%\end{equation}
%for $r^{(i)} \in [L_n-2 L_B, L_n]$, and zero otherwise.
\begin{equation}
\mathbf{B}^{(i)} = 
\begin{cases} 
     B_0 \sin \left( \frac{\pi (L_n-r^{(i)})}{2 L_B}   \right) \mathbf{\hat{r}}^{(i)} \times \mathbf{\hat{y}} \\
     \qquad \qquad \qquad \text{if $r^{(i)} \in [L_n-2L_B, L_n$],}  \\
   0   \qquad \qquad \qquad \hspace{-1ex} \text{otherwise.}
\end{cases}
\label{EqBInitial}
\end{equation}
Here $B_0$ is the initial strength of the magnetic field,
and $L_B$ the half-width of the initial ribbons. 
%The initial electric fields are consistent
%with the plasma $E\times B$ flow and the initial current
%densities are consistent with the magnetic field.  We choose
%$L_z = L_n$ and $L_z = 1.5 L_n$.

\begin{figure*}[t!]
\centering
\includegraphics{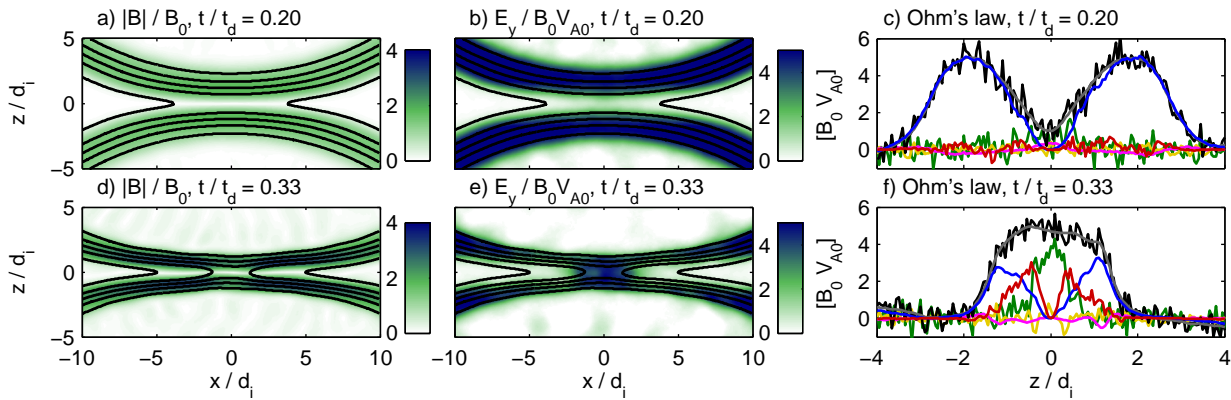}
\caption{(a,d) Magnetic fields $|B| / B_0$. (b,e) Reconnection
electric fields $E_y / V_{A0} B_0$.  (c,f) Generalized
Ohm's law $E_y = - (\mathbf{V_i} \times \mathbf{B})_y + (1/ne) (\mathbf{j} \times \mathbf{B})_y
-(1/ne) \partial P_{e,y} / \partial t - (1/ne) \nabla_\perp T_{e,y\perp}$.
Black: sum of Ohm's law RHS;
gray: $E_y$, blue: $-V_{i,z} B_x$ (with $V_i$ the ion flow);
red: $(1/ne) (j \times B)$;
green: $-(1/ne) \times \partial_x T_{e,yx}$;
yellow: $-(1/ne) \times \partial_z T_{e,yz}$;
pink: $-(1/ne) \partial P_{e,y}/\partial t$, where $P_{e,y}$ is the
momentum density in the electron fluid. 
The stress tensor $T_{e,ij}$ includes the inertia terms $m_e n v_{e,i} v_{e,j}$ due
to electron flow along the $i$ and $j$ directions.}
\label{FigPileupOhm}
\end{figure*}

Most parameters can be matched to the experiment, except
for a heavy electron mass and relatively low speed of light, as is typically
the case for PIC simulations.
Lengths are scaled to units of the ion-skin
depth to correctly retain the two-fluid effects compared to the experiments.
$L_n$ ranges from 20--100~$d_i$,
for the Rutherford and Omega experiments.
$L_B$ was found from experimental measurements and 
associated simulations \cite{LiPRL2007b} to be $\sim 0.1 L_n$ in Omega.  Note that
the current sheet width is therefore of order the ion skin depth, 
which is essential for the two-fluid reconnection mechanisms
discussed here.   $B_0$ is scaled to the plasma pressure,
to achieve the high-beta regime of the experiments.

Particle simulations were conducted with the 
relativistic, electromagnetic, explicit particle-in-cell
code \PSC \cite{Ruhl2006}.
We choose the PIC parameters
$M_i/m_e = 100$, $T_{i0} = T_{e0} = 0.02 m_ec^2$, and
an initial 100 particles per cell.
Total energy is conserved to better than 1\% in the simulations reported.
The results here are for collisionless simulations.
While this is largely consistent with the above estimate that 
reconnection in these experiments is dominated
by collisionless mechanisms, collisions are clearly also non-zero, and
this will be an important part of future modeling
as greater experimental fidelity is sought.

Figure \ref{FigBevol2d} presents the evolution of the
magnetic field in one simulation with 
parameters close to those of
the Rutherford experiments \cite{NilsonPRL2006},
$L_n = 20 d_i$, $n_b/n_0 = 0.1$, $L_B/L_n = 1/6$, 
$B_0^2 / n_0 T_{e0} = 0.25$, $V_0 = 3 (T_{e0}/m_i)^{1/2}$.
The image backgrounds show $|B|$, over which are plotted contours
of $\Psi$, which is defined
such that $\mathbf{B}(x,z) = \mathbf{\hat{y}} \times \nabla \Psi + B_y(x,z) \mathbf{\hat{y}}$.
Panel (a) gives the initial condition 
of Eq.~\ref{EqBInitial}.  The next three panels show
the evolution at later times, with time scaled in units of the 
dynamical time $t_d = L_n / V_0$.
As the bubbles interact, a reconnection x-line forms
at the leading point of tangency between the bubbles.
Near $t/t_d = 0.42$, the reconnection is about half complete,
and the reconnection rate is near its maximum.
Notice the strong pileup of
flux upstream of the reconnection region.
In contrast to reconnection considered in other contexts, 
here the strong inflows have sufficient
force to compress the flux in the current sheet by a ratio of about 4. 
This will be shown momentarily to have important
consequences for the scaling of the peak reconnection
rate.  Finally, by $t/t_d = 0.76$, in (d), reconnection is complete, and
the two bubbles have almost completely merged.
Notably, complete reconnection has occurred within about
1 dynamical time.

Figure~\ref{FigPileupOhm} shows, at two characteristic times 
(a,b,c: $t/t_d = 0.2$, (d,e,f): $t/t_d = 0.33$), $|B|$, the
reconnection rate $E_y = \partial \Psi / \partial t$, and 
a detailed cut of how the generalized Ohm's law
is fulfilled along a cut across the x-line.
The time $t/t_d = 0.2$ is characteristic of the
compression or pileup phase: the reconnection rate is small,
and the electric field is simply supported by
the $\mathbf{E} \times \mathbf{B}$ plasma flow of the ribbons (c).
However, by $t / t_d = 0.33$ the reconnection
is now proceeding at its maximum rate.
The electric field, of similar magnitude to that
earlier in the flow, is now sustained (1) in the current sheet by the 
Hall effect ($\mathbf{j} \times \mathbf{B} / ne$), since the current sheet
is now of the ion skin depth scale, and (2) at the x-line
by the the electron stress tensor.  Similar to typical results from
2-d PIC reconnection studies, within the stress tensor,
the off-diagonal pressure component $(\partial_x T_{yx})/ne$ 
is dominant, accounting for 80\% of the electric field.

%Figure~\ref{FigRecRate} shows the reconnection rate $E_y  = d \Psi/ dt$
%evaluated at the x-point.  (In this run, the current sheet did not
%tear into multiple islands, so there was only one x-point.)  
We find that the reconnection rate $\partial \Psi / \partial t$ 
sustained at this time is super-\Alfven{}ic compared to the 
\textit{nominal} $B_0 V_{A0}$, where $B_0$ is given 
in the initial condition,
and $V_{A0}$ is evaluated with $n_0$ and $B_0$.  
Furthermore, $E_y/B_0V_{A0}|_{max} \simeq 4.7$ is about a 
factor of 50 higher than the typical $0.1 B V_{A}$ found from
reconnection simulations including the Hall effect.
Here we have normalized to the nominal \Alfven{} speeds
to correspond to the experiments: the magnetic fields 
measured (and simulated), e.g. in Fig.~4(c1) of Ref.~\cite{LiPRL2007b},
are in fact an \textit{initial} condition before the bubbles have begun
to interact.

This point is important because two-fluid reconnection
mechanisms (the Hall effect and, in these PIC simulations, 
the electron pressure tensor), 
are typically found to allow for reconnection inflows of 0.1--0.2~$V_A$.
On its face, this is not sufficient to 
explain the apparent \Alfven{}ic reconnection observed in the experiments.
However, in this strongly driven regime, the \Alfven{} speed is also
time-dependent, due to the compression of the magnetic flux.  
Therefore we also track the \textit{instantaneous}
\Alfven{} speed using the maximum upstream
magnetic field $|B|$ and the density in the current sheet.
Indeed, once one accounts for the factor of 4 magnetic flux compression
(a factor a 16 in $B V_A$), and
the slightly reduced density at the current sheet
($n/n_0 \simeq 0.4$) one finds
that $E_y / B V_{A} \simeq 0.25$, much more
in line with typical two-fluid physics results.
Therefore, we find that previously-established two-fluid physics 
results can also apply to this geometry, once
one accounts for the flux pile-up.

\begin{figure}
\centering
\includegraphics{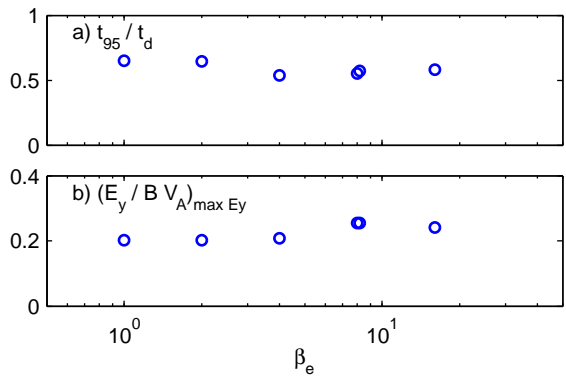}
\caption{(a)  Time for 
complete reconnection, and (b) peak reconnection rate
normalized to the instantaneous \Alfven{} speed,
vs. initial plasma $\beta_{e} = 2n_0T_{e0}/B_0^2$.}
\label{FigFluxPileup}
\end{figure}

Finally, we confirm the importance of flux-pileup in
this strongly driven regime.  Figure~\ref{FigFluxPileup} shows a sequence
of simulations in which we have varied the
strength of the initial magnetic field $B_0$, keeping
other parameters fixed.  The 
initial inflows become increasingly super-\Alfven{}ic
compared to the nominal \Alfven{} speed as
the initial magnetic field is decreased.  We then
measure the magnetic reconnection rates in two
ways.  First, in Fig.~\ref{FigFluxPileup}(a)
we measure the time for total reconnection,
based on the time in the simulation to reconnect
95\% of the initial flux [$\Psi_0 = (4/\pi) L_B B_0$].
Interestingly, we find that the 
total reconnection time only depends
quite weakly on $B_0$; therefore in the strongly
driven regime the reconnection time is insensitive
to the nominal \Alfven{} speed.

At the same time, we can also study the 
system at the time of peak reconnection.
Figure~\ref{FigFluxPileup}(b) shows
the peak reconnection rates, normalized
to the instantaneous \Alfven{} speeds,
calculated as above.  We find that with these corrections,
over a wide range of initial magnetic fields and
therefore flux pileup ratios,
the ultimate reconnection rates are consistently about
0.2 $B V_A$, roughly in line with previous two-fluid investigations.
%(We do not expect exact numerical agreement  because the
%current sheets are not quantitatively identical, e.g. to those
%produced out of a tearing mode.  The point is rather
%the dominance of the flux-pileup effect in controlling the reconnection rate.)

In past research, flux pileup has been found to operate
when reconnection is driven faster than allowed by the 
reconnection model (resistive MHD~\cite{BiskampPoF1986}, or
Hall~MHD~\cite{DorelliJGR2003}).
If the inflow rates are larger than the reconnection rates, 
then flux must pile up.
This is apparent in Fig.~\ref{FigPileupOhm},
where the reconnection rate at the x-line is less than the
flux advection rate from the plasma flow.  Meanwhile, the two-fluid physics result
$E_y \propto B V_A \propto B^2$ finds that 
reconnection rates increase sharply with pileup, 
so that eventually these two effects can come into
balance.  
%The question therefore becomes what
%can drive the flows to produce the pileup.  In this case,
%the source is the extraordinary plasma expansion driven by the
%lasers.
%Finally, 
However, pileup can only proceed if no other effects can
stop the compression process, and an important consideration 
here would be the accumulation of plasma and magnetic pressure
in the current sheet.  However, with the view of the laser-driven
experiments, where the initial current sheet densities
are much smaller than the bubble core densities, we
do not believe this would be sufficient to prevent reconnection.

In fact, our simulations find substantial
pressure increase in the current sheet, 
mostly due to strong heating of the current sheet
by compression.  The observation of current sheet heating is 
qualitatively consistent with the Rutherford experimental
results, which attributed the heating to 
magnetic energy dissipation \cite{NilsonPRL2006}.  
We suggest instead that the converging plasma flows far dominate 
the magnetic fields as an energy source, and that the heating is
from compression.

To conclude, recent experiments on magnetic
reconnection in laser-produced plasma bubbles
are found to be in a strongly driven reconnection
regime unexplored in previous laboratory experiments, 
where a combination of flux-pileup and
two-fluid effects account for extremely fast reconnection.
The pile-up is found to be an enormous effect, 
boosting the relevant \Alfven{} speeds to
match the specified plasma inflow.  The strong
inflow and flux pileup lead to a reconnection time 
independent of the nominal \Alfven{} time.

This research is supported by the DOE Grant No. DE-FG02-07ER46372 
and the NSF.  We thank Drs.\ R. Petrasso, C.-K. Li, and F.\ S\'{e}guin for
useful discussions, and Prof.\ H.\ Ruhl for initially providing the
\PSC{} code.  A. B. acknowledges gratefully the support of a 
Fulbright Scholar award.
% which has been adapted for this problem by W.F. and K.G.

\bibliographystyle{apsrev}

%\bibliography{citeulike,extra} 

\end{document}